\documentclass[journal]{IEEEtran}
\usepackage{amsmath}
\usepackage{amssymb}
\usepackage{graphicx}
\usepackage{color}
\usepackage{cite}
\usepackage{caption}

\ifCLASSINFOpdf
\else
\fi
\begin{document}
	\bstctlcite{IEEEexample:BSTcontrol}
%	\makeatletter
%	\def\bstctlcite{\@ifnextchar[{\@bstctlcite}{\@bstctl
%			cite[@auxout]}}
%	\def\bstctlcite[#1]#2{\@bsphack
%		\@for\@citeb:=#2\do{%
%			\edef\@citeb{\expandafter\@firstofone\@citeb}%
%			\if@filesw\immediate\write\csname #1\endcsname{\s
%				tring\citation{\@citeb}}\fi}%
%		\@esphack}
%	\makeatother
%\title{\Large Reconciliation Throughput Analysis of the Satellite-based QKD}
\title{Optimised Multithreaded CV-QKD Reconciliation for Global Quantum Networks}
\author{
	\IEEEauthorblockN{Xiaoyu Ai and Robert Malaney}\\
	\IEEEauthorblockA{School of Electrical Engineering  \& Telecommunications,\\
		University of New South Wales, Sydney, NSW 2052, Australia.}\\

}

\vspace{-3cm}

\maketitle


\begin{abstract}		
Designing a practical Continuous Variable (CV) Quantum Key Distribution (QKD) system requires an estimation of the quantum channel characteristics and the extraction of secure key bits based on a large number of distributed quantum signals. Meeting this requirement in short timescales is difficult. On standard processors, it can take several hours to reconcile the required number of quantum signals. This problem is exacerbated in the context of Low Earth Orbit (LEO) satellite CV-QKD, in which the satellite flyover time is constrained to be less than a few minutes. A potential solution to this problem is massive parallelisation of the classical reconciliation process in which a large-code block is subdivided into many shorter blocks for individual decoding. However, the penalty of this procedure on the important final secured key rate is non-trivial to determine and hitherto has not been formally analysed. Ideally, a determination of the optimal reduced block size, maximising the final key rate, would be forthcoming in such an analysis. In this work, we fill this important knowledge gap via detailed analyses and experimental verification of a CV-QKD sliced reconciliation
protocol that uses large block-length low-density parity-check decoders. Our new solution results in a significant increase in the final key rate relative to non-optimised reconciliation. In addition, it allows for the acquisition of quantum secured messages between terrestrial stations and LEO satellites within a flyover timescale even using off-the-shelf processors. Our work points the way to optimised global quantum networks secured via fundamental physics.
\end{abstract}

%\begin{keywords}
%	\textit{\textbf{Index Terms--}}\textbf{Quantum Key Distribution (QKD), LDPC Codes, Key Reconciliation, Satellite Communications}
%\end{keywords}

% \fi
\IEEEpeerreviewmaketitle
{\color{black}
\section{Introduction}\label{Introduction}

Continuous Variable (CV) Quantum Key Distribution (QKD) has been intensively studied and significant breakthroughs have been achieved in both theory and experiment (see \cite{hosseinidehaj2018satellite} for review). Compared to Discrete Variable (DV) QKD \cite{bennett1984update,weinfurter2016quantum,gyongyosi2019survey,pirandola2020advances}, CV-QKD can be implemented with well-developed technologies (e.g., homodyne detectors) in commercial fibre-optic networks\cite{korzh2015provably,eriksson2019wavelength} and free-space optical communications\cite{shen2019free,gyongyosi2019secret}, providing it a potential advantage in practical deployments\cite{jouguet2011long,jouguet2012field,liao2018long,guo2020comprehensive,zhang2020long}.

Considering the finite-key security of CV-QKD and DV\nobreakdash-QKD, there are three critical parameters. These are, $N_o$, the number of original quantum signals sent by the transmitter (Alice) that are collected by the receiver (Bob); $N_e$, the number of quantum signals from which the protocol parameters are estimated;\footnote{More precisely, in a CV-QKD protocol, Alice and Bob randomly select a $N_e$-signal subset from the $N_o$ signals to estimate the parameters.} and,  $\epsilon$, the probability that a QKD protocol fails to generate secret keys\cite{leverrier2010finite,furrer2012continuous}. To satisfy an upper limit on the failure probability of parameter estimation, Alice and Bob   set  $N_e$ to a large value, which in turn implies a larger $N_o$.

Despite the advantages in deployment, CV-QKD systems tend to demand a larger $N_o$ to reach the same $\epsilon$ relative to DV-QKD protocols. For example, to achieve a final key rate of $0.1$ bits per pulse with $\epsilon=10^{-9}$, a CV-QKD protocol studied in \cite{kish2020feasibility} required $N_o\approx 10^{9}$ signals. However, to achieve the same final key rate with $\epsilon = 10^{-14}$, the DV-QKD protocol in \cite{tomamichel2012tight} required $N_o\approx 10^4$ signals. This higher number of required signals in CV-QKD can render the classical post-processing (i.e. key reconciliation and privacy amplification\footnote{In this work, we focus on the key reconciliation step because it is the more time-consuming part in the post-processing steps while the privacy amplification involving only bit-wise operations can be easily implemented faster than the reconciliation~\cite{yuan201810}.}) slow -  possibly failing to meet target timescales for reconciliation.

The end-users of a CV-QKD system expect the system to deliver two identical and secure keys under a limited time interval. For example, for satellite-based deployments, we would hope that the reconciliation is completed while maintaining a line-of-sight connection with the ground station. For a CV-QKD-enabled satellite with orbital parameters similar to \emph{Micius} \cite{liao2017satellite}, this would mean the reconciliation should be completed in less than a few minutes. For the protocol we use in this work (see later), and for $\epsilon=10^{-9}$, this, in turn, would require the data rate of reconciliation to be at least $3.6 \times 10^6$ bits per second. For real-time reconciliation (say in sub-second timescales), two orders of magnitude increases in the reconciliation rates would be required. Demands for smaller $\epsilon$ will exacerbate the issue. Ideally, the rate of reconciliation should always be faster than the rate of quantum signalling.

This all raises the question as to whether current CV-QKD reconciliation schemes are  optimised for the highest possible key rates in bits per second. As we show here, this is  not the case. Further optimisation is possible on all current schemes.

To understand the issue better, we define reconciliation in the context of CV-QKD as a two-step scheme where the inputs to the reconciliation are non-identical $N=2N_o - 2N_e$ quadrature values\footnote{ $N_o$ and $N_e$ are multiplied by 2 since Alice and Bob utilise both quadratures  from  heterodyne detection - the detection process we assume in this work.} held by Alice and Bob (after parameter estimation), and the output is an identical bit string held by Alice and Bob \cite{lin2015high,wang2017efficient,zhou2019continuous}. Assuming a reverse reconciliation scheme, Bob first converts the quadrature values encoded by Alice in each signal   to $m$ bits. Alice, after converting each of her encoded real numbers also to $m$ bits, then initiates some discrepancy-correction algorithms based on pre-defined error-correction codes to ensure her $mN$ bits are identical to Bob's. In this work, we will adopt Low-Density Parity-Check (LDPC) codes for the error correction.

However, as alluded to above, reconciling $mN$ bits within a limited time frame can be challenging. State-of-the-art LDPC-based reconciliation schemes for CV-QKD systems involve parallelised computation on a Graphics Processing Unit (GPU) \cite{milicevic2017key,guo2020comprehensive} or Field-Programmable Gate Arrays (FPGAs)\cite{yang2020high,li2021fpga}. Reconciliation schemes implemented on FPGAs offer more programmable flexibility, but sometimes at the cost of reduced memory access relative to GPUs. For our purposes, both hardware architectures are useful - both offer massive parallelisation opportunities. These parallelisation solutions generally take the following two-step approach: 1) The $mN$ bits are organised as $m$ $N$-bit blocks to be reconciled. Each $N$-bit block is divided into multiple shorter blocks of size, say, $N_R$. This is usually just set to a block size that can be processed within some timescale. 2) Then the $m$ $N_R$-bit blocks are reconciled in parallel (via independent processors) using optimally-designed LDPC decoders. However, what is missing in this approach is a proper optimisation analysis as to what the optimal value of $N_R$ is. As we show below, simply reducing $N_R$ at the cost of additional processing units is not an optimal solution. It transpires that in QKD the ``penalty'' cost of reducing the code rate (implicit in the use of small block lengths) significantly influences the bit per second final key rate.

A more sophisticated analysis is required  to determine the optimal reduced block length. Such an analysis is the key contribution of this work. Although we will adopt a specific CV-QKD protocol for our analysis, the key steps of our scheme will apply to any CV-QKD protocol. Our reconciliation scheme will deliver the highest reconciliation  rate for a given processor speed - thus allowing for the optimal solution to CV-QKD reconciliation.

\section{System Overview}
\label{section:SystemOverview}
Although, as just stated, our analysis will apply to most CV-QKD protocols, for detailed quantitative discussion we will consider only one specific CV-QKD protocol - the ``no-switching'' protocol\cite{weedbrook2004quantum,hosseinidehaj2020finite,dequal2020feasibility} based on heterodyne detection. In this protocol, the quantum signal is encoded using Gaussian-modulated coherent states\cite{weedbrook2004quantum}. The main advantage of the no-switching protocol is that Alice and Bob can utlise all  measurement results \cite{hosseinidehaj2020finite} (in most other protocols some results are discarded due to a random quadrature selection). We also adopt a Slice Reconciliation (SR) variant named Multi-Stage Hard Decoding\footnote{The slice reconciliation can be implemented with 2 other variants: Bit Interleaved Coded Modulation (BICM)~\cite{bloch2005efficient} and Multi-Level Coding/Multi-Stage Decoding (MLC/MSD)~\cite{jouguet2014high,mani2021multiedge}. We note that the MLC/MSD takes advantage of the dependence between slices to select the optimal LDPC code rates~\cite{bloch2005efficient,mani2021multiedge}. However, as a special case of MLC/MSD, MSHD assumes that the slices are independent~\cite{imai1977new,wachsmann1999multilevel,bloch2005efficient}. Using MSHD leads to a tractable analysis at the expense of sub-optimal selection of LDPC code rates, but such expense is negligible if Gray Labelling~\cite{wesel2001constellation} is adopted and the number of slices is at least 5~\cite{wachsmann1999multilevel,bloch2005efficient}, as is the case in this work.} (MSHD)~\cite{imai1977new,wachsmann1999multilevel,bloch2005efficient} for the classical reconciliation step, where the number of bits derived from each measurement outcome is $m$. We refer to this variant simply as SR in the following.

It is worth noting that the optimisation analysis to follow is to some extent independent of the details of the reconciliation scheme. However,
SR\cite{bloch2005efficient,bloch2006ldpc} can be compared with the other well-known reconciliation scheme for CV-QKD - multi-dimensional reconciliation\cite{leverrier2008multidimensional}. It is known that SR achieves higher reconciliation efficiency when the Signal-to-Noise Ratio (SNR) is greater than 1\cite{bloch2006ldpc}. At low SNR the opposite is true.
For focus, here we adopt SR (as multistage hard decoding\cite{bloch2006ldpc})  since in many satellite scenarios post-selection is used to filter out the low SNR quantum signals
 \cite{hosseinidehaj2018satellite}. Our adopted scheme will be more useful in such scenarios.

\begin{figure*}[h]
	\centering
	\includegraphics[width=\textwidth]{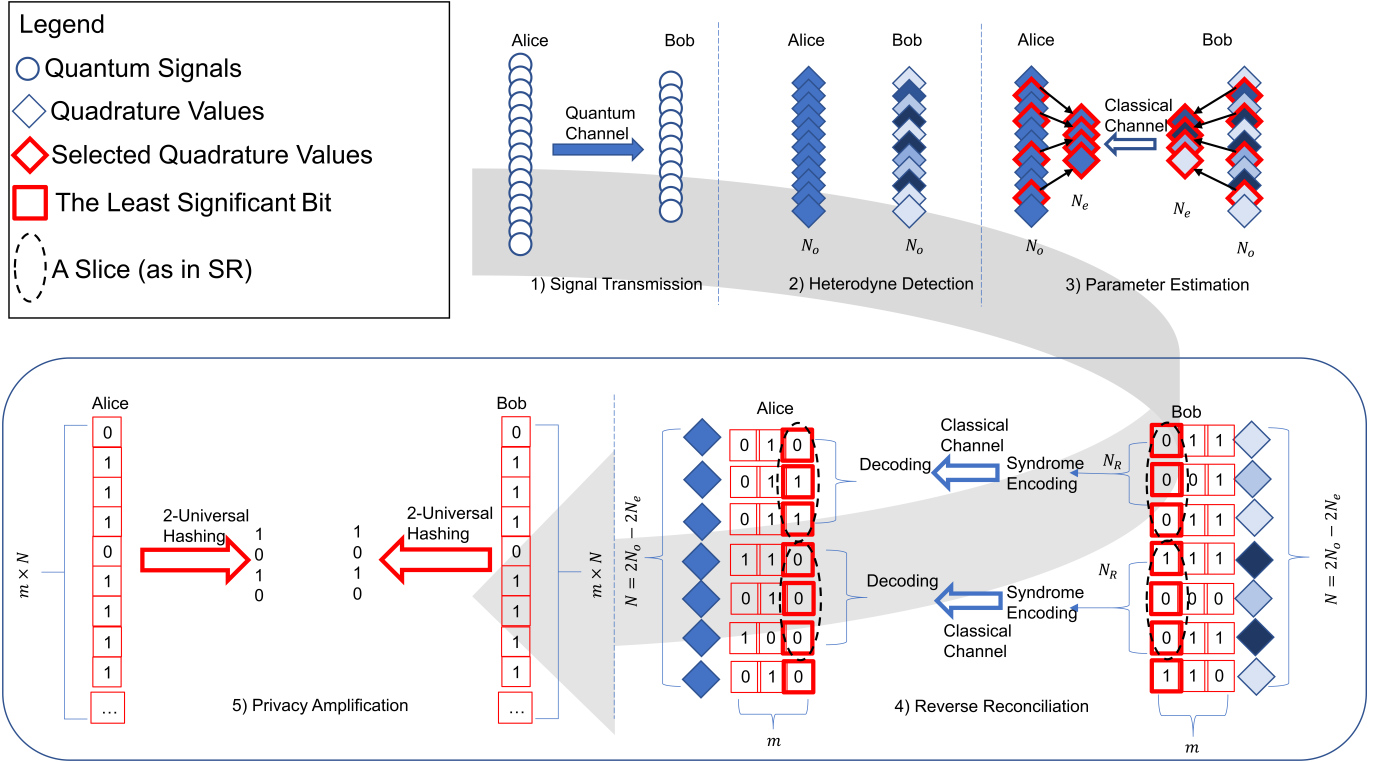}
	\caption{Diagram of the reconciliation. 1) Alice prepares and sends Gaussian-modulated coherent states to Bob during the signal transmission. 2) Bob performs heterodyne detection to obtain quadrature values. Blue diamonds at Alice's side represent Alice's quadrature values. Diamonds filled with tinted or shaded blue at Bob's side show that his quadrature values deviate from what was prepared by Alice after transmission. 3) Alice and Bob select and exchange a random subset of the quadrature values (diamonds with red outlines) to perform parameter estimation. 4) Alice and Bob perform reverse reconciliation based on SR to convert their quadrature values into two bit strings and reconcile them. 5) Alice and Bob perform privacy amplification to obtain two identical but shorter key strings about which Eve has effectively no knowledge.\label{fig:NDiagram}}	
\end{figure*}

We now briefly describe the steps of the protocol, a diagram of which shown in Fig.~\ref{fig:NDiagram}. In the following, we assume the quantum signalling rate is much larger than the reconciliation rate.
\begin{itemize}
	\item \textbf{Step 1: Signal Preparation.} Alice selects a fixed modulation variance $V_A$. For each quantum signal to be transmitted to Bob, Alice randomly selects a number from a Gaussian distribution, $N(0,V_A)$, and then prepares a signal by displacing one of the quadrature components  of a vacuum state by this random number. The process is repeated on the signal for the other quadrature. The signal is then transmitted to Bob.
	
	\item \textbf{Step 2: Heterodyne Detection.} Bob performs heterodyne detection to obtain the two quadrature values (real numbers) for each received signal. Bob compares each measured quantum signal with a given cut-off threshold and informs Alice to discard her corresponding quantum signal if his measured quantum signal is lower than the threshold\footnote{The Gaussian post-selection technique at Bob's side effectively improves the channel conditions between Alice and Bob\cite{fiuravsek2012gaussian} so that SR is preferred for reconciliation (rather than multidimensional reconciliation).}. A quantum signal that is lost in transit registers a null signal at Bob. Neglecting null signals, Bob holds $2N_o$   quadrature values at the end of this process.

	\item \textbf{Step 3: Parameter Estimation.} Bob randomly selects a subset $2N_e$ from the $2N_o$ quadrature values and sends this \emph{estimation subset}, along with the corresponding time information, to Alice via classical communications (we adopt $N_e = \frac{1}{2}N_o$, unless otherwise stated).
 Alice uses the timing information to best pair the signals in this subset (and therefore the corresponding quadrature values) sent by her and then estimates the covariance matrix between the shared states. Based on the estimated covariance matrix, Alice determines the channel transmissivity, $T$, excess noise, $\xi$,  Bob's SNR, $\gamma$, the Holevo Information, $\chi_{BE}$, between Bob and the eavesdropper (Eve), and the mutual information between Alice and Bob, $I_{AB}$. Finally, for a given target reconciliation efficiency $\beta$, Alice compares $\chi_{BE}$ with $\beta I_{AB}$. Alice aborts the protocol if $\chi_{BE} \geq\beta I_{AB}$. Otherwise, Alice informs Bob of the estimation results, i.e. $T$, $\xi$, $\gamma$, $\chi_{BE}$ and $I_{AB}$.

	\item \textbf{Step 4: Bit Error Estimation for SR.} Using Gray Labelling, Alice and Bob represent each of the quadrature values embedded in each signal with $m$ bits. Then, for quadrature values selected in the estimation subset, Alice forms a $N_{e}$-by-$m$ bit matrix and Bob does the same. Next, Alice and Bob exchange their matrices and compare the $j^{th}$ column of the two matrices to estimate the Bit Error Ratio\footnote{At this step, sources of bit errors include the channel transmission, heterodyne detection, and quantisation.} (BER), $p_j,j\in \{0,1,\cdots,m-1\}$, for all the digits in the $j^{th}$ column. The estimated $p_j$ will be used in SR. Finally, Alice and Bob discard all the quadrature values in the estimation subset. At the end of this step, Alice and Bob each hold a $mN$-bit string.
	
	\item \textbf{Step 5: Reverse Reconciliation.} For each column, Alice and Bob agree on an LDPC code with block length $N_R$ that is closest to the capacity determined by $p_j$. Bob forms a new $N_R$-bit string (referred as a ``slice'' in SR) by selecting the $j^{th}$ digit (bit) of each of the $N_R$ quadrature values, encodes the new bit string (the slice) into syndrome bits, and sends those bits to Alice (see III.B for details). Alice then initiates SR to obtain her best estimate of Bob's string. Alice repeats this process until all her $mN$ bits are reconciled.
%The detailed steps of the SR will be explained in Section \ref{section:KeyReconciliation}.
Finally, Alice and Bob obtain two hashed strings by applying the same hash function to their reconciled strings and exchange the hash results to check whether SR is successful. If successful, Alice holds a $mN$-bit string  identical to Bob's $mN$-bit string. Otherwise, they abort the protocol and restart from Step 1.
	
	\item \textbf{Step 6: Privacy Amplification.} Based on Eq.~\ref{eq:BPSKeyRate}, Alice and Bob compute the length of the secret key that can be extracted and then apply a 2-universal hashing function on their reconciled string to obtain two identical and shorter secret key strings about which Eve has effectively no knowledge.
\end{itemize}

\label{section:KeyReconciliation}

\section{Overcoming the Limitations of Key Reconciliation}
\label{section:penalty}
%Recalling that the reconciliaton for quadrature values from  $p$ and $q$ quadrature measurements are the same, we focus our discussion and analysis to the quadrature values from one of the quadrature.
\subsection{GPU-based SR}
\label{section:DecodingTime}
The process of SR is to reconcile $mN$ bits. One can naively use $m$ LDPC  matrices with $N_R=N$ for each matrix. However, due to practical hardware limitations, the process is better implemented by dividing $N$ into $N_d$ blocks of some smaller $N_R$ so that the same LDPC decoders can reconcile these blocks in parallel. This process resembles the idea of \textit{Single Program Multiple Data} (see \cite{darema2001spmd,pharr2012ispc} for more details). As illustrated in Fig.~\ref{fig:GPUSlice}, we implement SR by creating $N_d$ LDPC decoders loaded with the same LDPC  matrix on $N_d$ GPU threads and let these decoders reconcile $N_d$ blocks in parallel. This helps to reduce the SR timescale and assists in meeting the time constraints, such as those posed in satellite-based scenarios. Section \ref{section:TimeSimulation} will demonstrate in detail the advantage of using such parallelisation.

\begin{figure}[h]
	\centering
	\includegraphics[width=0.5\textwidth]{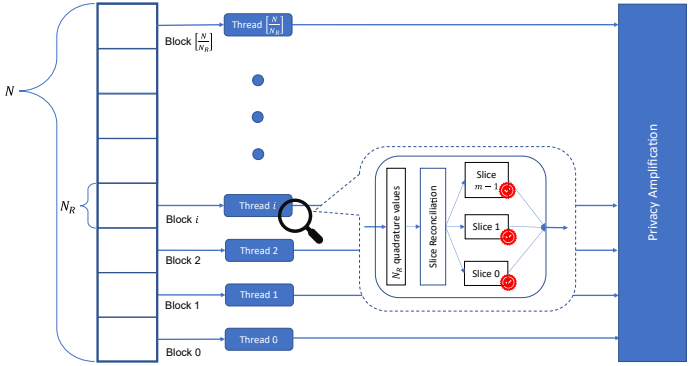}
	\caption{Diagram of GPU-based SR adopted in this work. $N$ quadratures values are divided into multiple blocks of length $N_R$. Each block is loaded to one thread that is dedicated to performing SR for this block. The reconciled bits obtained from each thread are collected and ready for privacy amplification.  The floor operator is shown as $[ \cdot ]$. \label{fig:GPUSlice}}
\end{figure}

\subsection{The Penalty of Using Finite-Length LDPC Codes}
%In our SR scheme, we assume that Alice and Bob adopt reverse reconciliation and Alice runs LDPC decoders.
An illustration of the SR scheme is shown in Fig.~\ref{fig:SliceDiagram}. The generic steps are: \emph{1)} for the $i^{th}$ quadrature value, $y^i, i=0,1,\cdots,N_R-1$, Bob applies a constant-step quantisation function, $M(\cdot)$, to convert $y^i$ to an $m$-bit string\footnote{We assume the least significant bit is $l_0^i$.} denoted as $\{l_0^i, l_1^i, \cdots, l_{m-1}^i\}$, where $l_j^i, j=0, \cdots, m-1$ is the binary bit for the $j^{th}$ digit of the $i^{th}$ quadrature value. \emph{2)}. We define that the $j^{th}$ slice, $\mathbf{S_j}$, is a bit string with length $N_R$ created by Bob: $\mathbf{S_j} = \{l_j^0, l_j^1, \cdots, l_j^{N_R-1} \}$. For $\mathbf{S_j}$, Bob applies an LDPC matrix, $H_j$ based on $p_j$ obtained in parameter estimation to obtain the corresponding syndrome bits. \emph{3)} Bob sends Alice the syndrome bits of $\mathbf{S_j}$ and $H_j$ via classical communications. \emph{4)} Alice uses her quadrature values as side information and what was transmitted by Bob as the inputs of the LDPC decoder. Alice takes the soft decoding output (the log-likelihood ratio when the decoding finishes) of $\mathbf{S_{j-1}}$ as the input to accelerate the reconciliation of $\mathbf{S_j}$ (except for $\mathbf{S_0}$)\footnote{The rationale behind this is that the soft decoding output of $\mathbf{S_{j-1}}$ provides \textit{a priori} information on the reliability of each bit in $\mathbf{S_{j}}$ \cite{bloch2005efficient}.}\cite{bloch2005efficient,lodewyck2007quantum}. \emph{5)} Alice obtains her estimated version of $\mathbf{S_{j}}$. Then, Alice and Bob move on to $\mathbf{S_{j+1}}$. \emph{6)} Alice and Bob repeat Step 1 to 5 until all $N$ values are reconciled. We note that Alice and Bob use $m$ LDPC matrices to reconcile $m$ slices in a block  - but the same $m$ LDPC matrices are used for reconciling all $N_d$ blocks since the quantisation errors are the same for a given $m$\cite{laudenbach2018continuous}.

\begin{figure*}[h]
	\centering
	\includegraphics[width=0.8\textwidth]{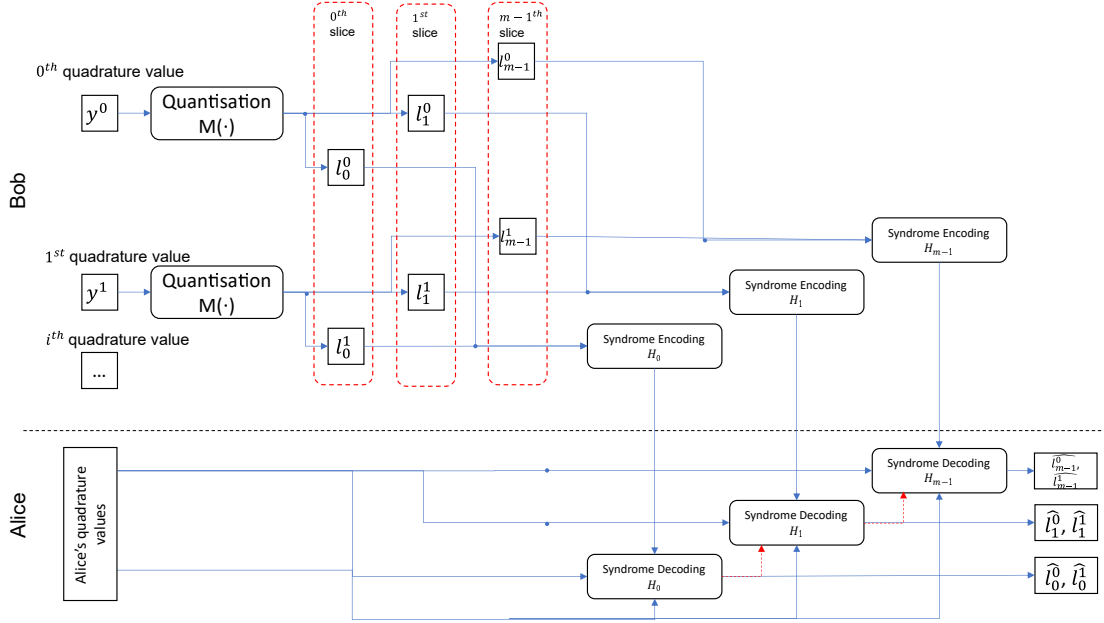}
	\caption{Diagram of SR. The red dashed arrows show the soft decoding output feeds from $\mathbf{S_{j-1}}$ to $\mathbf{S_j}$. The red dashed rectangles are graphical examples of a slice. \label{fig:SliceDiagram}}
\end{figure*}

In SR, Bob needs to transmit syndrome bits to Alice based on the selected LDPC  matrix with the code rate, $R_j$, for $\mathbf{S_j}$ via classical communications. For a given channel condition, selecting $R_j$ closest to the capacity is the common approach to minimise the number of bits disclosed to the eavesdropper while Alice can still reconstruct Bob's quantised bits without error\cite{bloch2006ldpc}. Specifically, for a given $T$, we can obtain the SNR, $\gamma$, as \cite{laudenbach2018continuous}
\begin{equation}
\label{eq: SNR}
\gamma=\frac{\frac{1}{2}V_AT}{1+\frac{1}{2}\xi}\,,
\end{equation}
where $V_A$ is the modulation variance at Alice side,  $\xi = \xi_{ch} + \xi_d$ is the total noise power, $\xi_{ch}$ is the channel excess noise, and $\xi_d$ is the detector noise. Finally, the reconciliation efficiency $\beta\in [0,1]$ for the SR is obtained via \cite{jouguet2014high}
\begin{equation}
\label{equation: beta}
\beta = \frac{\Pi(M(Y))-m+\sum_{j=0}^{m-1}R_j}{I_{AB}}\,,
\end{equation}
where $Y$ is a vector of Bob's quadrature values of length $N_R$, $M(Y)$ is a $mN$-bit string obtained by applying the quantisation function $M(\cdot)$ to each quadrature value in $Y$, and $\Pi(M(Y))$ is the entropy function of $M(Y)$. Increasing $m$ to values that render the quantisation error negligible is always possible, but this would require the individual LDPC codes for every $j^{th}$ slice to be near perfect (capacity-achieving) otherwise the efficiency $\beta$ will be low; $m=5$ is found to be a good pragmatic compromise, and is adopted here. Given five slices a constant quantisation size of the real line across $2^5$ bins centered on zero is chosen. This size, which is dependent on the adopted $\gamma$,  optimises $\beta$  (see\cite{jouguet2014high} for further discussion).

%In \cite{jouguet2014high}, $\beta$ is decomposed to two parts: the quantisation efficiency and the code efficiency. In SR, increasing $m$ improves $\beta$ by increasing the former efficiency close to one, assuming capacity achieving codes are used for each slice. However, setting $m>5$ is unnecessary for a SNR ranging from $0.5$ to $10$ because the information loss due to quantisation is less than $1\%$\cite{jouguet2014high} when $m=5$. Therefore, we will neglect the information loss due to quantisation in our analysis later.

%For a given channel, $\beta$ is dependent on the $m$-bit quantisation process and the choice of $R_j$ for each slice. Following the same approach in \cite{jouguet2014high}, we rewrite Eq.~\ref{equation: beta} as
%\begin{equation}
%\label{eq:betaRewrite}
%\begin{aligned}
%\beta =& \frac{H(M(Y))-m+\sum_{j=0}^{m-1}R_j}{I_{AB}}\\
%=&\left( \left(H(M(Y))-m + \sum_{j=0}^{m-1}C^{Binary}_j \right) \right.\\
% & \left.+ \left( \sum_{j=0}^{m-1}\beta_cC^{Binary}_j - \sum_{j=0}^{m-1}C^{Binary}_j\right)\right) \\
%=&\frac{H(M(Y))-m + \sum_{j=0}^{m-1}C^{Binary}_j}{I_{AB}} \\
%&- \left(1-\beta_c\right)\frac{\sum_{j=0}^{m-1}C^{Binary}_j}{I_{AB}}\,.\\
%\end{aligned}
%\end{equation}
%where $C^{Binary}_j$ is the binary channel capacity for $\mathbf{S_{j}}$ and $\beta_c = R_j/C^{Binary}_j$.

The LDPC code rates, $R_j$, in Eq.~\ref{equation: beta} are the actual rates of the specific codes used for each slice (of length $N_R$). Normally, in practice, $N_R$ is simply set to some value that allows  target time-frames to be met, given that the decoding time is an increasing function of the block length\cite{milicevic2017key}. We use $R_j$ to obtain our experimental key rate in Eq.~\ref{eq:ExpKeyRate}. A more nuanced value of that $N_R$ that optimises secure key rates is now analysed.

To make progress in our task, we utilise a previous analysis of channel coding in the finite block-length\cite{polyanskiy2010channel} regime as a means to further investigate the effective channel capacity, $C_{Finite}$ for a given block length $N_R$ and $\gamma$. For a finite message set $\mathcal{M}$, $C_{Finite}$ is the ratio of the maximum size of $\mathcal{M}$ that can be transmitted via $N_R$ channel uses with a decoding error probability less than $\epsilon_{EC}$. Specifically, for an Additive White Gaussian Noise Channel (AWGNC), $C_{Finite}$ is given by\footnote{This approximation is accurate if the code achieves more than $80\%$ of the capacity\cite{polyanskiy2010channel}. } \cite{polyanskiy2010channel}
\begin{equation}
\label{eq:R_Finite}
C_{Finite} \approx C(\gamma)-\frac{\sqrt{N_RA}Q^{-1}(\epsilon_{EC}) + \frac{1}{2}\log N_R}{N_R}\,,
\end{equation}
where $C(\gamma) = \frac{1}{2}\log (1+\gamma)$ is the Shannon Capacity for the given $\gamma$, $Q^{-1}(s)$ is the inverse of the Q-function
\begin{equation}
	Q(z)=\frac{1}{\sqrt{2\pi}}\int_{z}^{\infty}e^{-\frac{t^2}{2}}dt\,,
\end{equation}
 and $A$ is given by
\begin{equation}
A = \frac{\gamma}{2} \frac{\gamma+2}{(\gamma+1)^2}(\log \epsilon_{EC})^2\,.
\end{equation}
Function $A$ is termed the ``channel dispersion'' since it represents the reduction of the code rate from the channel capacity due to a tolerated decoding error probability. It is  the ``price to pay" for using a code with finite block length, for a given $\gamma$.

Note, $C_{Finite}$ is the upper bound of $\sum_{j=0}^{m-1}R_j$ for a given $\epsilon_{EC}$ and $N_R$. To simplify the determination of the code rate in the finite-length regime, we determine $C_{Finite}$ instead of each $R_j$ for the purpose of analysis. Using Eq.~\ref{eq:R_Finite} we introduce $\beta_{Finite}$  as an analytical reconciliation efficiency in the finite LDPC block length regime (neglecting the information loss due to the quantisation process). This is given by
\begin{equation}
\label{eq: BetaFinite}
\beta_{Finite} = \frac{C_{Finite}}{I_{AB}}\approx \frac{{I_{AB}}-\frac{\sqrt{N_RA}Q^{-1}(\epsilon_{EC}) + \frac{1}{2}\log N_R}{N_R}}{I_{AB}}\,.
\end{equation}
Eq.~\ref{eq: BetaFinite} explicitly illustrates how LDPC codes with long block lengths generally reduce the information disclosed to Eve during reconciliation.

We demonstrate the connection between $\beta_{Finite}$ and $\beta$. Firstly, we rewrite Eq.~\ref{equation: beta} as 
	\begin{equation}
	\label{equation: beta2}
	\beta = \frac{\Pi(M(Y))-R_s}{I_{AB}}\,,
	\end{equation}
	where $R_s = \sum_{j=0}^{m-1}(1 - R_j) = m -\sum_{j=0}^{m-1}R_j$ is the ratio of syndrome bits sent by Bob to the total number of bits in $m$ slices. $R_s$ is the side-information that Alice uses to reconcile her $m$ slices\cite{mani2021multiedge}. It is known that $R_s$ satisfies the Slepian\nobreakdash Wolf Bound~\cite{slepian1973noiseless}
	\begin{equation}
	\label{equation: SWB}
	R_s \geq \Pi(M(Y)|X)\,,
	\end{equation}
	where $X$ is a vector of Alice's quadrature values of length $N_R$. Applying Eq.~\ref{equation: SWB} to Eq.~\ref{equation: beta2}, we have 
	\begin{equation}
	\begin{aligned}
	\frac{\Pi(M(Y))-R_s}{I_{AB}} &\geq  \\
	&\frac{\Pi(M(Y))-\Pi(M(Y)|X)}{I_{AB}} =  \frac{I(M(Y);X)}{I_{AB}}\,,	
	\end{aligned}	
	\end{equation}
	where $I(M(Y);X)$ is the total mutual information (after quantisation) between Alice and Bob. Recalling that $ C_{Finite} $ is the upperbound of the mutual information between Alice and Bob for an LDPC block length, we have the following
	\begin{equation}
	1 \geq \beta  \geq \frac{I(M(Y);X)}{I_{AB}} \geq \frac{C_{Finite}}{I_{AB}} = \beta_{Finite}\geq 0\,.
	\end{equation}

%We also note that $C(\gamma)$ is $I_{AB}$ because the information shared between Alice and Bob reduces to classical information after the heterodyne detection.

\subsection{Analysing the Computational Complexity of SR}
An LDPC  matrix with block length $N_R$ can be defined by the symbol and check node degree distribution polynomials, $\lambda(x)=\sum_{a=2}^{\Lambda}\lambda_a x^{a-1}$ and $\rho(x)=\sum_{b=2}^{P}\rho_b x^{b-1}$. Here, $\Lambda$ and $P$ are the highest degrees in $\lambda(x)$ and $\rho(x)$, respectively. We denote the total number of non-zero entries in an LDPC matrix as $G$, and adopt the well-known Belief Propagation (BP) decoder\cite{richardson2008modern} for error correction. We define the total number of arithmetic operations of SR as $\sum_{j=0}^{m-1} E_j  D_{j}$, where, for each $\mathbf{S_j}$, $E_j$ is the number of arithmetic operations executed within a decoding iteration,\footnote{In a BP decoder, a decoding iteration is one pass through the decoding algorithm.} and $D_{j}$ is the number of decoding iterations\cite{ai2020reconciliation}. We note, in our GPU-based SR, $E_j$ and $D_j$ are different for the $m$ slices of each block since $m$ LDPC matrices are used to reconcile the $m$ slices. For a channel with constant $T$ and $\xi$, $D_{j}$ is dependent on a target $\epsilon_{EC}$, and on the polynomials $\lambda(x)$ and $\rho(x)$. Note, for $N_R$  larger than approximately  $10^5$, $D_j$ is independent of $N_R$ (a result we will adopt later). Assuming the  Gaussian approximation within the Density Evolution Algorithm, $D_j$ is given by
\begin{equation}
\label{eq:rob2}
	D_{j} = \arg \min_{k} \{ q_k = f(\gamma, k, \lambda(x), \rho(x)) \leq \epsilon_{EC},k \in \mathbb{Z^*}\}\,,
\end{equation}
where $q_k$ is the BER after the $k^{th}$ decoding iteration and given by\cite{chung2001analysis}
%Due to the assumption that the channel between Alice and Bob is a BI-AWGNC, we adopt the Gaussian Approximation (GA) approach to implement $ f(\Gamma, k, \lambda(x), \rho(x))$ and calculate $D_{i}$ in our work. It is shown in \cite{chung2001analysis} that DEA with GA connects the degree distribution of an LDPC code, decoding error and decoding iterations. To obtain $D_{i}$, we use the following iterative equation\cite{chung2001analysis}
\begin{equation}
\label{eq:rob}
\begin{aligned}
q_k &= f(\gamma, k, \lambda(x), \rho(x))\\
&=\sum_{b=2}^{P} \rho_b \phi^{-1}\left(1 - L^{b-1} \right)\,.
\end{aligned}	
\end{equation}
Here
\begin{equation}
	L = 1 - \sum_{a=2}^{\Lambda}\lambda_a \phi\left(\log \gamma+\left(a-1\right)q_{k-1}\right)\,,
\end{equation}
where $q_0=0$, and $\phi(v)$ is given by
\begin{equation}
\label{eq:phiFunc}
\phi(v) = \begin{cases}
 1 - \frac{1}{\sqrt{4\pi v}}\int_{-\infty}^{+\infty} \tanh\left(\frac{u}{2}\right)e^{-\frac{\left(u - v\right)^2}{4v}}du & v>0\\
 1& v=0\,.
\end{cases}
\end{equation}
%We can numerically calculate the integral in Eq.~\ref{eq:phiFunc}. However, this is
Finding a closed solution to Eq.~\ref{eq:rob} is problematic due to the $\phi^{-1}(w)$ term (here $w=\phi(v)$). To make progress, the following approximation for Eq.~\ref{eq:phiFunc} is used\cite{chung2001analysis}
\begin{equation}
	\phi(v) \approx \begin{cases}
		e^{-0.4527v^{0.86}+0.0218} & v>0\\
	1& v=0\,.
	\end{cases}
\end{equation}
We then find $\phi^{-1}(w)$ is given by
\begin{equation}
	\phi^{-1}(w) \approx \begin{cases}
	\left( \frac{\log{w}- 0.0218}{-0.4527}  \right)^{1.1628} & 0<w<1\\
	0& w=1\,.
	\end{cases}
\end{equation}
With this all in place, it is now possible to solve for $D_j$ as given by Eq.~\ref{eq:rob2}.
%where $p_0 = Q(2S)$ is the initial bit error probability and can be converted using the estimated channel SNR and $f(p_0, k, \lambda(x), \rho(x))$ is the density evolution function for the decoder used in our paper:
%\begin{equation}
%\label{EqDEGeneral}
%\begin{split}
%p_k=&\int_{-\infty}^{0} \Omega_p^{0} dx = \int_{-\infty}^{0}  \left\lbrace P_{L|C}(u,j) \Omega_p^{2u}+(1-P_{L|C}(u,j))  \right.\\
%&\left.  \Omega_p^{2u} *  \lambda(\Gamma^{-1}(\rho(\Gamma(\Omega_p^{2u+2})))) \right\rbrace dx
%\,,\end{split}
%\end{equation}
%and
%\begin{equation}
%\Omega_p^{2u}=p_0\delta(x-\log \frac{p_0}{1-p_0})+(1-p_0)\delta(x+\log \frac{p_0}{1-p_0})+(1-p_0))
%\end{equation}

Now we focus on the determination of $E_j$. When messages are propagated from the variable nodes to the check nodes, there are $2G$ multiplications and $G$ additions\cite{chandrasetty2011fpga}. When messages are propagating back to the variable nodes, there are $4G$ operations required ($2G$ multiplications and $2G$ additions)\cite{chandrasetty2011fpga}. Therefore, $E_j$ is obtained by\cite{chandrasetty2011fpga,ai2020reconciliation}
\begin{equation}
\label{eq: EP}
\begin{aligned}
E_j &=7G\\
&= 7N_R (\frac{\sum_{b=2}^{P}\frac{\rho_b}{b}}{\sum_{a=2}^{\Lambda}\frac{\lambda_a}{a}})(\sum_{b=2}^{P}b\rho_b)\,.
\end{aligned}	
\end{equation}
The decoding time of the whole reconciliation process, $\Delta t$, is given by
\begin{equation}
\label{eq:DeltaT}
	\Delta t = c_h  \sum_{j=0}^{m-1}E_j  D_{j}\,,
\end{equation}
where $c_h$ is a hardware-dependent constant representing the time taken to complete an arithmetic operation. Clearly, by dividing $N$ values into multiple blocks with length $N_R$ and decoding these blocks simultaneously, Alice and Bob can reduce the decoding time by a factor of $N_d = \frac{N}{N_R}$.

\section{Final Key Rate}
\label{section:KeyRate}
We now present the penalty incurred for the division of $N$ in the finite-key regime, and then propose an optimisation procedure to find the optimal $N_R$ which maximises the final key rate in bits per second.
%In this section, our analysis and discussions are based on quadrature values obtained from both the quadrature measurement.
\subsection{Analysis of the Final Key Rate}
\label{section:Analysis}
For the protocol considered, in the finite-key regime the final key rate in bits per pulse, $K$, under the assumption of Gaussian collective attacks is adopted as \cite{leverrier2015composable,lupo2018continuous,hosseinidehaj2020finite}\footnote{This key-rate formulation   was developed in \cite{leverrier2015composable} with a typographical error corrected in \cite{lupo2018continuous}. Eq.~\ref{eq:BPSKeyRate} is from \cite{hosseinidehaj2020finite} which acknowledged the correction and simplified the final key rate formulation (see footnote 2 of \cite{hosseinidehaj2019optimal} for more details). A general discussion on the use of other key-rate formulations (e.g. \cite{pirandola2021limits,pirandola2021satellite}) within our framework is given later.}
\begin{equation}
\label{eq:BPSKeyRate}
K=\frac{N(\beta I_{AB} - S_{BE}^{\epsilon_{PE}} ) - \sqrt{N}\Delta_{AEP}-2\log_2\frac{1}{2\epsilon_{PA}}}{N_o}\,,
\end{equation}
where $S_{BE}^{\epsilon_{PE}}$ is the upper bound of the estimated $\chi_{BE}$ (note $K$ is an upper bound, which we assume is reached).
%Strictly speaking, $K$ is upper bounded by the RHS of Eq.~\ref{eq:BPSKeyRate}. To simplify our calculation, we assume $K$ reaches this upper bound. We assume the same for $K_{Finite}$ and $K_{Finite}^\prime$ in the rest of this work.}
The determination  of $S_{BE}^{\epsilon_{PE}}$ is carried out and utilised in the key rates derived here, but this determination is somewhat lengthy. As such, the reader is referred to the appendix for a full explanation and derivation of this term. We simply note here that $S_{BE}^{\epsilon_{PE}}$ is dependent on estimates of the channel parameters and therefore on the value of $N_e$, the number of symbols sacrificed in the estimation.
In Eq.~\ref{eq:BPSKeyRate}, $\Delta_{AEP}$ is a penalty term (derived using the Asymptotic Equipartition Property of a stochastic source) due to the finite number of bits used in quantisation and privacy amplification, and is given by
\begin{equation}
\label{eq:AEP}
\begin{aligned}
\Delta_{AEP} =& (m+1)^2 + 4(m+1)\sqrt{\log_2 (\frac{2}{\epsilon^2_{s}})} \\
&+2\log_2(\frac{2}{\epsilon^2\epsilon_{s}}) + \frac{4\epsilon_{s}m}{\epsilon\sqrt{N}}\,,
\end{aligned}	
\end{equation}
where $\epsilon = \epsilon_{EC} + 2\epsilon_{s} + \epsilon_{PA}+\epsilon_{PE}$ is the probability that a QKD protocol fails to generate secret keys. Here, $\epsilon_{s}$ is the smoothing parameter associated with the smooth min-entropy calculation,  $\epsilon_{PA}$ is the failure probability of the privacy amplification, and $\epsilon_{PE}$ is the probability that the true value of $\chi_{BE}$ is not within the confidence interval calculated during parameter estimation. For a given $\epsilon$, one can determine the values of $\epsilon_{EC}$, $\epsilon_{s}$, $\epsilon_{PA}$, and $\epsilon_{PE}$ by setting them individually or collectively as part of the maximisation of $K$ (see Eqs.~18 -- 21 in \cite{kish2020feasibility} for details).

We consider the penalty on the final key rate caused by dividing $N$ into sub-blocks of length $N_R$. Replacing $\beta$ in Eq.~\ref{eq:BPSKeyRate} with the $\beta_{Finite}$ of Eq.~\ref{eq: BetaFinite}, we obtain the final key rate with the finite LDPC block length effect fully considered. The new rate is given by
\begin{equation}
\label{eq:FiniteK}
K_{Finite} = \frac{N(C_{Finite} - S_{BE}^{\epsilon_{PE}} ) - \sqrt{N}\Delta_{AEP}-2\log_2\frac{1}{2\epsilon_{PA}}}{N_o}\,.
\end{equation}

Thus far, we have been analysing the final key rate in bits per pulse. However, the final key rate in bits per second is more interesting in our context - the system will not be viable if the reconciliation takes too long to complete. From this point forward, we use a dashed symbol to distinguish a final key rate that is given in bits per second. Taking the decoding complexity into account, and ignoring the time taken to acquire the quantum signals, we can write the final key rate, $K_{Finite}^\prime$, as
\begin{equation}
\label{eq:BPSRate}
\begin{aligned}
K_{Finite}^\prime &= \frac{N_o K_{Finite}}{\Delta t}\\
& = \frac{N(C_{Finite} - S_{BE}^{\epsilon_{PE}} ) - \sqrt{N}\Delta_{AEP}-2\log_2\frac{1}{2\epsilon_{PA}}}{c_h\sum_{j=0}^{m-1}E_j  D_{j}}\,.
\end{aligned}
\end{equation}
We observe that $\Delta t$ in Eq.~\ref{eq:DeltaT} and $C_{Finite}$ in Eq.~\ref{eq:R_Finite} are increasing functions of $N_R$. We are interested in finding a unique $N_R$ so that $K_{Finite}^\prime$ is maximised.

\subsection{Optimised LDPC Blocklength for CV-QKD Reconciliation}
\label{section:Optimal}
%In the last two sections, we know that applying block-wise paralellisation can reduce the decoding time with the expanse of increased information disclosure to Eve. We are interested in investigating whether there is an optimal $N_R$ to achieve the maximal $K_{Finite}\prime$.

%As shown in Eq.~\ref{eq:BPSRate}, for a given $\epsilon_{EC}$, $\epsilon_{s}$, $\epsilon_{PA}$, and $\epsilon_{PE}$, both the numerator and the denominator of $K_{Finite}^\prime$ are increasing functions of $N_R$.
Previously, we have shown that parallelisation reduces the decoding time at the expense of increased information disclosure to Eve. In this section, we demonstrate an optimisation process to find the unique $N_R$ maximising $K_{Finite}^\prime$.

We consider a scenario where $\epsilon_{EC}$, $\epsilon_{s}$, $\epsilon_{PA}$, and $\epsilon_{PE}$ are manually set by end users. We can formulate the optimisation problem for the scenario
\begin{equation}
\label{eq:Optimisation1}
\begin{aligned}
\max_{N_R} \quad & K_{Finite}^\prime\\
\textrm{s.t.} \quad & 10^5\leq N_R \leq N \,,\\
\end{aligned}
\end{equation}
where $K_{Finite}^\prime$ is defined in Eq.~\ref{eq:BPSRate}. The lower limit of $10^5$ arises from our earlier discussion on ensuring $D_j$ is independent of $N_R$ (for smaller values of $N_R$ the penalty cost will be prohibitive and we ignore this range). We notice that $\Delta t$ is a linear function of $N_R$\footnote{Recalling Eq.~\ref{eq:DeltaT}, we note that for a given LDPC code, $E_j$ is only dependent on the degree distribution pairs and $D_{j}$ is only a function of the degree distribution pairs and $\gamma$.} and  $S_{BE}^{\epsilon_{PE}}$; and that $\Delta_{AEP}$ and $\epsilon$ are independent of $N_R$. Therefore, we can rewrite Eq.~\ref{eq:Optimisation1} in the simplified form
\begin{equation}
\label{eq:simplifiedOpt}
\begin{aligned}
\max_{N_R} \quad & \frac{N C_{Finite} - B_1}{B_2 N_R}\\
\textrm{s.t.} \quad & 10^5\leq N_R \leq N\,,\\
\end{aligned}
\end{equation}
where
\begin{eqnarray}
B_1=&N S_{BE}^{\epsilon_{PE}}  + \sqrt{N}\Delta_{AEP}+2\log_2\frac{1}{2\epsilon_{PA}}\,\label{eq:B1}\,,\\
B_2=&\sum_{j=0}^{m-1}7(\frac{\sum_{b=2}^{P}\frac{\rho_b}{b}}{\sum_{a=2}^{\Lambda}\frac{\lambda_a}{a}})(\sum_{b=2}^{P}b\rho_b)  D_{j}c_h\,.
\end{eqnarray}

To solve this optimisation problem, firstly we show that the second derivative of $K_{Finite}^\prime$ with respect to $N_R$ is less than zero for all $N_R$ considered, where
\begin{equation}
\label{eq:2diff_eq}
\begin{aligned}
\frac{d^2 K_{Finite}^\prime}{dN^2_R} =& \frac{-1}{4B_2 N^4_R}\left(\left(12N \log N_R -10 N  \right) \right.\\
&\left. + 15 \sqrt{A}Q^{-1}(\epsilon_{EC}) N\sqrt{N_R} \right.\\
&\left. - \left( 8I_{AB}N N_R - 8B_1 N_R\right)   \right) \,.
\end{aligned}	
\end{equation}
To show that the RHS of Eq.~\ref{eq:2diff_eq} is less than zero for all $N_R$ considered, it is equivalent to show
\begin{equation}
\label{eq:2diff_eq2}
\begin{aligned}
12N \log N_R -10 N + 15 \sqrt{A}Q^{-1}(\epsilon_{EC}) N\sqrt{N_R}\\
> \left( 8I_{AB}N N_R - 8B_1 N_R\right)   \,.
\end{aligned}	
\end{equation}
We find that the LHS of inequality~\ref{eq:2diff_eq2} is greater than zero for $N>N_R>10$, $\gamma>0$, and $\epsilon_{EC}<\frac{1}{2}$, but the sign of the RHS is subject to specific CV-QKD parameters. However, through detailed numerical search we find that the inequality~\ref{eq:2diff_eq2} holds for the range of parameters anticipated for realistic CV-QKD deployments.\footnote{In using this technique it is important to check that the inequality holds for the chosen parameter range of interest. This is done for all calculations we show here, but also for a much wider range not shown. For example, we find for
 $\epsilon=10^{-9}$, $N=10^9$ and $m=5$, the inequality~\ref{eq:2diff_eq2} holds for any combination of the remaining parameters selected from the ranges  $V_A\in[1,34]$, $T\in[0,1]$ and $\xi=[0,0.05]$. }
% We also found that Inequality~\ref{eq:2diff_eq2} holds for any $V_A\in[1,51]$, $T\in[0,0.1]$ and $\xi=[0,0.05]$, given $\epsilon=10^{-10}$, $N=10^9$ and $m=5$.}
%\footnote{To reduce the parameter search space, we conducted the numerical search for some given $\epsilon$, $N$ and $m$ that are commonly considered in satellite-based scenarios. Given $\epsilon=10^{-9}$, $N=10^9$ and $m=5$, we found that inequality~\ref{eq:2diff_eq2} holds for any $V_A\in[1,30]$, $T\in[0,0.92]$ and $\xi=[0,0.2]$. We also found that Inequality~\ref{eq:2diff_eq2} holds for any $V_A\in[1,45]$, $T\in[0,0.94]$ and $\xi=[0,0.4]$, given $\epsilon=10^{-10}$, $N=10^9$ and $m=5$.}
%\footnote{Based on our numerical search, inequality~\ref{eq:2diff_eq2} holds for  any combination of the parameters provided the range of each parameter satisfies the following. $N\in \left[10^6, 10^{11}\right]$, $m\in \left[1,10\right]$, $\epsilon\in \left[10^{-8}, 10^{-12}\right]$, $V_A \in \left[1,20\right]$, $T\in \left[0.1, 0.93\right]$, $\xi \in [0,0.3]$.}
%\footnote{Based on our numerical search, inequality~\ref{eq:2diff_eq2} holds for the following ranges for each parameter. $N\in \left[10^7, 10^9\right]$, $m\in \left[3,6\right]$, $\epsilon\in \left[10^{-8}, 10^{-10}\right]$, $V_A \in \left[3,7\right]$, $T\in \left[0.7, 0.93\right]$, $\xi_{ch}\in \left[0.01, 0.05\right]$ and $\xi_d \in \left[0.01, 0.05\right]$.}
For example, if we consider the following  CV-QKD settings\footnote{The values of $\xi_{ch}$ and $\xi_d$ in the standard CV-QKD settings are predicted values after accounting for all noise terms \cite{kish2020feasibility}.} (in the following we refer to these as the standard CV-QKD settings); $N=10^9$, $m=5$, $\epsilon_{EC} = 2\epsilon_{s} = \epsilon_{PA} = \epsilon_{PE} = 2.5\times10^{-10}$, $V_A = 5$, $T = 0.9$, $\xi_{ch}=0.0186$ and $\xi_d=0.0133$, we find that the LHS of the inequality is greater than $10^{12}$ and the RHS of the inequality is less than $10^{11}$.

To find the maximised $K_{Finite}^\prime$, we first find the value of $N_R$ that satisfies $\frac{d K_{Finite}^\prime}{d N_R} = 0$, where
\begin{equation}
\label{eq:diff_eq}
\begin{aligned}
\frac{d K_{Finite}^\prime}{d N_R} & = \frac{N_R N \frac{dC_{Finite}}{dN_R}-\left(N C_{Finite} - B_1\right)}{B_2 N^2_R}\\
&=-\frac{NI_{AB}}{B_2 N^2_R} - \frac{N}{2B_2 N^3_R}+\frac{3N\sqrt{A}Q^{-1}(\epsilon_{EC})}{2B_2 N^{\frac{5}{2}}_R } \\
&+\frac{N\log N_R}{B_2 N^3_R} + \frac{B_1}{B_2 N^2_R}\,. \\
%&=-\frac{NI_{AB}}{N^2_R B_2} - \frac{N-2N\log N_R}{2B_2 N^3_R}   \\
%&+\frac{3N\sqrt{A}Q^{-1}(\epsilon_{EC})}{2N^{\frac{5}{2}}_R B_2} + O(N^{-3}_R)\,.
\end{aligned}	
\end{equation}
Therefore, our equation to be solved is given by
\begin{equation}
\label{eq:diff_eq2}
-\frac{NI_{AB}}{B_2 N^2_R} - \frac{N-2N\log N_R}{2B_2 N^3_R}  +\frac{3N\sqrt{A}Q^{-1}(\epsilon_{EC})}{2B_2 N^{\frac{5}{2}}_R } + \frac{B_1}{B_2 N^2_R}= 0\,.
\end{equation}
Eq.~\ref{eq:diff_eq2} can be solved via a numerical root-finding algorithm\cite{mathews2004numerical}. If we consider the
standard CV-QKD settings, we obtain a stationary point at $N_R = 3.6\times 10^7$ bits (the value of $K^\prime_{Finite}$ at this $N_R$ is discussed later).
%However, in cases where other CV-QKD settings are used, multiple $N_R$ or even no $N_R$ can be found by solving $\frac{d K_{Finite}^\prime}{d N_R} = 0$ because $\frac{d^2 K_{Finite}^\prime}{dN^2_R}$ is not necessarily less than zero for the $N_R$ considered. In those cases, further constraints will be added to the optimisation process to find the unique optimal $N_R$.

In closing this section we note the following in regard to alternate key-rate equations. Although we have adopted a specific CV-QKD protocol and specific key rate equation, different security analyses of our adopted protocol, and analyses of different  CV-QKD protocols, can have a key rate with a similar form to that shown in Eq.~\ref{eq:BPSKeyRate} albeit with different bounded rates. When comparing different key-rate equations it is perhaps useful to only consider the leading terms. This allows for clearer tractability in determining the optimal $N_R$. For example, if the last term in our Eq.~\ref{eq:AEP} is neglected (a good approximation for reasonable $N$ values) then the functional dependence on $N$ of many key-rate equations is identical. In such circumstances, the same key rate can be mapped to alternate security settings of the different key-rate equations, and our framework applies directly. For example, using the key rate equation of \cite{pirandola2021limits} we find the same optimal $N_R$ albeit with a normalised key rate of one for the following security settings (described with the notations in\cite{pirandola2021limits}): the number of quantum signal exchanged, $N=10^9$, the number of quantum signals sacrificed for parameter estimation, $m=\frac{N}{2}$, the probability of successful reconciliation, $p_{ec} = 0.95$, the smoothing parameter, $\epsilon_{s} = 3.2\times10^{-13}$, the hashing parameter $\epsilon_{h} = 4.7\times10^{-13}$, the probability that the true value of the square-root transmissivity is less than the value obtained by the worst-case estimator used in parameter estimation, $\epsilon_{pe} = 1.2\times 10^{-13}$, the residue probability that Alice and Bob's bit strings are different after passing the error correction, $\epsilon_{cor} = 0.8\times 10^{-13}$, the modulation variance, $\sigma^2_x = 7.1$, the size of the effective alphabet after quantising the continuous quadrature values, $d=2^5=32$, the channel transmissivity, $T = 0.81$, the channel noise $\sigma^2_z = 0.035$, and the quantum duty to pay by the detector, $\nu_{det}=2$ for heterodyne detection. Key-rate equations with different functional dependence on $N_R$ can still be analysed within the framework proposed here - albeit via modified optimisation relations. Examples of this arise in consideration of DV-QKD protocols (albeit for which reconciliation optimisation is usually less important). We also note that extension of our adopted CV-QKD key-rate equation, to cover the most general attack, is possible via the use of the Gaussian de Finetti reduction technique and the inclusion of an energy test \cite{leverrier2017security}. This leads to a scaling of order $N^4$ in terms of the security cost.

\section{Experimental Results}
\label{section:Experiment}
We conducted an experiment of our GPU-based SR on a NVIDIA GTX 1060 GPU (with 6GB GPU memory). The GPU provides up to 1280 Compute Unified Device Architecture (CUDA) threads that can be run simultaneously. We determine the BER after decoding, denoted as $p_{Decode}$ (different from the $p_j$ obtained at Step 5 of the protocol). We also measured the decoding time for $mN$ bits. Below, we determine the experimental final key rate, $K_{exp}^\prime$, and compare it with $K_{Finite}^\prime$ to verify the optimality of $N_R$ found in Section \ref{section:Optimal}. We note that the experimental data shown in Figs.~\ref{fig:BER},~\ref{fig:DTime}, and~\ref{fig:KeyRateNR} is averaged over 50 runs. We also note, in our specific GPU the number of threads available was less than the number of blocks when $N_R=10^5$. This was numerically compensated for in the results shown.

In the experiment, we assume that Alice and Bob complete the first four steps of the protocol described in Section~\ref{section:SystemOverview}. Since Alice and Bob's quadrature values are the input and output of an AWGNC, we can generate these quadrature values for SR in the following way. 1) For a given $V_A$, $T$, and $\xi$, we obtain the noise power, $\sigma^2_n$ of the AWGNC from Eq.~\ref{eq: SNR}
\begin{equation}
	\gamma = \frac{\frac{1}{2}V_AT}{1+\frac{1}{2}\xi} = \frac{1}{\frac{1+\frac{1}{2}\xi}{\frac{1}{2}V_AT}} = \frac{1}{\sigma^2_n}\,.
\end{equation}
2) We generate $N$ random numbers from the distribution $N(0, V_A)$. These numbers are regarded as Alice's quadrature values and denoted as $\mathbf{x} = \{x_0, x_1, \cdots, x_{N-1}\}$. 3) We obtain Bob's quadrature values $\mathbf{y}$, where the \emph{i}th component is given by $y_i = x_i + n $, and  where $n$ is a random real number drawn from $N(0,\sigma^2_n)$.
%These generated quadrature values are used to produce all the figures in this work.

\subsection{Decoding Error and Time}
\label{section:TimeSimulation}
In Fig.~\ref{fig:BER}, we compare the BER performance of different $N_R$ settings for each $T$ considered\footnote{Recall, we are particularly interested in the satellite-to-Earth channel. As in other works, we assume losses for this channel are dominated by diffraction effects, and therefore the transmissivity can be held constant. We further assume post-selections, using a bright classical beam sent along with the quantum signals (but different polarisation), remove any significant transmissivity deviations. As discussed elsewhere\cite{kish2020feasibility}, some receiver/transmitter apertures, coupled to detailed phase-screen simulations of satellite downlink channels, render the constant-transmissivity assumption reasonable\cite{villasenor2020atmospheric}. If the transmissivity is highly variable the optimal block length, $N_R$, can be calculated by an expectation over the transmissivity density function.}. The solid lines represent the best straight-line fit of the experimental data for each $N_R$. We note that for a given $T$, the LDPC code rates are set to $10\%$ lower than the capacity for that $T$. In Fig.~\ref{fig:BER}, we observe that larger LDPC codes lead to a lower $p_{Decode}$. This observation is consistent with the result of the finite-length information theory\cite{polyanskiy2010channel}. %Therefore, using a larger block length code leads to a $\beta$ closer to 1.

In Fig.~\ref{fig:DTime}, we determine the measured decoding time for $mN$ bits in the experiment, $\Delta t_{exp}$, which is normalised to the value at $N_R=10^9$ ($\Delta t_{exp}=310$ seconds). The differences of the measured decoding time at each $N_R$ reflect additional decoding iterations. Our results confirm the reduction of $\Delta t_{exp}$ when a smaller $N_R$ is used. We note that it is difficult to quantify the exact relation between $\Delta t_{exp}$ and $N_R$ since $\Delta t_{exp}$ includes the elapsed time taken by SR's arithmetic operations  and the elapsed time for the overhead; mostly due to memory access operations and synchronisation (we estimate this shortly).
%For example, the overhead can take up to $20\%$ of $\Delta t_{exp}$ for $N_R\geq10^6$.
However, our experiment generally confirms the advantage of parallelisation in decoding time. %Our results show the expected advantages and disadvantages of the block-wise parallelisation in our analysis.

\begin{figure}[h]
	\centering
	\includegraphics[width=0.5\textwidth]{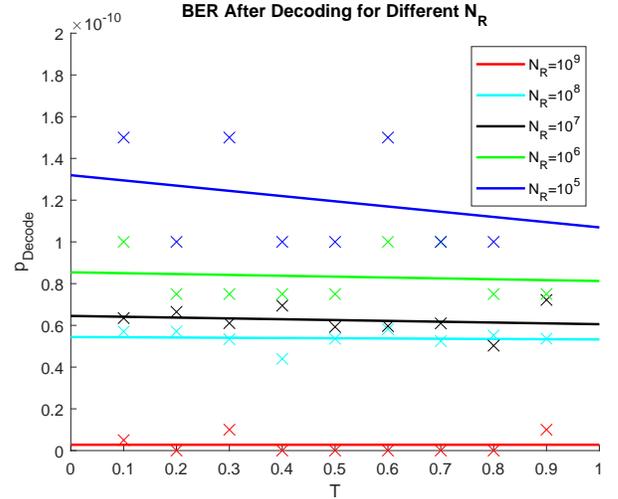}
	\caption{$p_{Decode}$ after decoding for different $T$ and $N_R$. For each colour, the crosses are the experimental data obtained.\label{fig:BER}}
\end{figure}
\begin{figure}[h]
	\centering
	\includegraphics[width=0.5\textwidth]{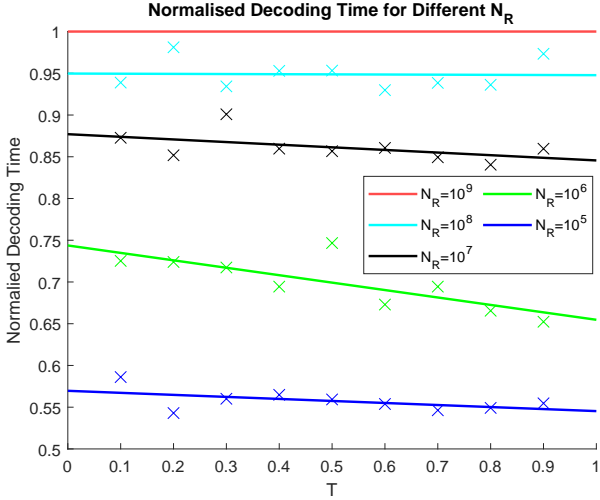}
	\caption{Normalised decoding time vs. $T$ for different $N_R$. For each colour, the  crosses are $\Delta t_{exp}$ obtained in our experiment. For each $T$, the data presented is the ratio of $\Delta t_{exp}$  for each $N_R$ to the decoding time for $N_R=10^9$.
%The times shown represent the total time taken to decode a block of length $10^9$ using parallel processing of smaller sub-blocks.
The advantage offered by  GPU parallel processing is seen to improve decoding times by $\sim 30\%$ for sub-blocks of length $10^6$.\label{fig:DTime}}
\end{figure}
%Comparing Figs.~\ref{fig:BER} and \ref{fig:DTime}, we observe that $\Delta t_{exp}$ for $N_R=10^5$ are not significantly reduced compared to the one with the setting $N_R=10^5$. In terms of the final key rate, we find no significant reduction when switching from $N_R=10^6$ to $N_R=10^5$, although the penalty of setting $N_R=10^5$ is observed. Our observation suggests that setting $N_R=10^5$ may also be a viable option for block-wise paralellisation in situations where the on-board memories are limited.

\subsection{The Optimal $N_R$}
\label{section:NRResult}
Previously, we have analytically found the optimal $N_R$ which maximises $K_{Finite}^\prime$. Now we wish to check the compatibility of this $N_R$ with the value that maximises $K_{exp}^\prime$ based on realistic LDPC codes and the specific GPU used in our experiment.

For this experiment, we pre-built a database to store LDPC codes with their code rates ranging from $0.01$ to $0.8$ and their block lengths ranging from $10^6$ to $10^9$. For code rates less than $0.1$, Multi-Edge-Type LDPC codes (degree distributions outlined in \cite{wang2017efficient}) were used. These achieve a lower $p_{Decode}$ compared to irregular LDPC codes with the same code rate. For code rates greater than or equal to $0.1$, we adopted the irregular LDPC codes (degree distributions outlined in\cite{mateo2011efficient}). At such code rates, these latter codes have the same $p_{Decode}$ performance as the Multi-Edge-Type counterparts, but allow for faster code construction. 

We use the following process to obtain $K_{exp}^\prime$ for each $N_R$ considered. 1) For each $\mathbf{S_{j}}$, we select an LDPC code whose code rate is closest to the capacity determined by $p_j$ from the pre-built database. 2) We use the selected code to test whether the probability of an error correction failure of that code is less than $\epsilon_{EC}$. 3) If the test fails, we decrease $R_j$ by $\Delta R = 0.05$, select another LDPC code from the database and go back to Step 2). Otherwise, we mark the selected code as ``good'' and then go back to Step 1) for the next slice. The process terminates when all slices have been successfully decoded. We then obtain $K_{exp}^\prime$ using
\begin{equation}
\label{eq:ExpKeyRate}
	K_{exp}^\prime = \frac{N(\sum_{j=0}^{m-1}R_j - S_{BE}^{\epsilon_{PE}} ) - \sqrt{N}\Delta_{AEP}-2\log_2\frac{1}{2\epsilon_{PA}}}{\Delta t_{exp}}\,.
\end{equation}

We note that the overhead mentioned earlier is one of the sources causing the discrepancy between $K_{exp}^\prime$ and $K_{Finite}^\prime$. To compensate the additional decoding time due to the overhead for all $N_R$ considered, we adopt a numerical search for a compensated $\Delta t_{exp}$ so that $|K_{exp}^\prime - K_{Finite}^\prime|^2$ is minimised. Our result shows that $K_{exp}^\prime$ after compensation is approximately $10\%$ higher than the uncompensated $K_{exp}^\prime$. In Fig.~\ref{fig:KeyRateNR}, we plot $K_{Finite}^\prime$ and $K_{exp}^\prime$ (compensated and uncompensated)  with respect to $N_R$ based on Eqs.~\ref{eq:BPSRate} and \ref{eq:ExpKeyRate}, respectively.

\begin{figure}[h]
	\centering
	\includegraphics[width=0.5\textwidth]{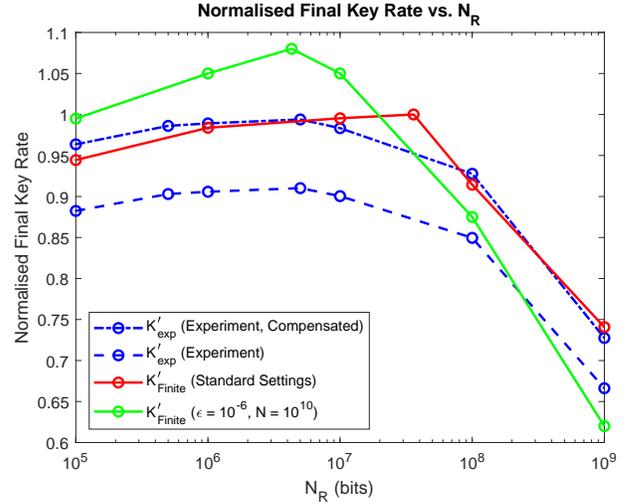}
	\caption{Final Key Rate vs. LDPC Block Length. Here, we adopt the standard CV-QKD settings for the red and the blue curves. For the green curve, we adopt the following settings: $N = 10^{10}$, $m=5$, $\epsilon_{EC} = 2\epsilon_{s} = \epsilon_{PA} = \epsilon_{PE} = 2.5\times10^{-7}$, $V_A = 4$, $T = 0.92$, $\xi_{ch}=0.0180$ and $\xi_d=0.0128$. We also set $c_h = 3.2\times 10^{-9}$ seconds for the red and green curves. All the curves are normalised to the maximum of the red curve, i.e. $K_{Finite}^\prime = 4.3 \times 10^5$ bits per second at $N_R = 3.6 \times 10^7$. \label{fig:KeyRateNR}}
\end{figure}

The optimal $N_R = 3.6 \times 10^7$ and $5\times 10^6$ are found for $K_{Finite}^\prime$ and $K_{exp}^\prime$, respectively. Assuming the usual practical scheme where $N_R$ is simply selected randomly, our results for the standard CV-QKD settings show that using the optimal $N_R$ for SR leads to a maximum gain of $33\%$ on the final key rate. Other CV-QKD settings will provide different maximum gains. For example, the green curve of Fig.~\ref{fig:KeyRateNR} provides for a $66\%$ gain (not shown in the figure are the rates for $N_R>10^9$). This point emphasises the need to consider the parameter settings before determining both the optimal $N_R$ and the gain achieved relative to the standard practice of simply picking some $N_R<N$. Note that in our rate determinations, the normalisation of one in Fig.~\ref{fig:KeyRateNR} corresponds to a key rate  $K_{Finite}^\prime = 4.3\times 10^5$ bits per second, based on our hardware-specific value of $c_h=3.2 \times 10^{-9}$ seconds.\footnote{We adopted the following method to determine $c_h$. For $m$ LDPC codes with $N_R = 10^6$ and $T=0.9$, we obtained the total number of arithmetic operations for those codes. Next, we measured the elapsed time to reconcile a block of $10^6$ quadrature values. We then obtained $c_h$ by dividing the number of arithmetic operations to this measured elapsed time.} The reconciliation rate associated with this same key rate is $3.9 \times 10^6$ bits per second. Assuming the source rate of the quantum signalling  was high enough (\emph{e.g.} a 100MHz source),
this reconciliation rate is higher than that required for delivery of secured ($\epsilon=10^{-9}$) keys ($N=10^{9}$) within flyover times (270 seconds) consistent with Micius-type orbits (see the introduction).

Comparing the two curves in Fig.~\ref{fig:KeyRateNR}, we find that there is still a small discrepancy between $K_{Finite}^\prime$ and $K_{exp}^\prime $ although they share a similar trend. The reason for such discrepancy is twofold.
Firstly, there remain small trapping sets in the LDPC matrices.\footnote{These trapping sets are the primary reason that additional decoding iterations are consumed for only a marginal decrease of the decoding error, i.e. the error floor effect\cite{tian2004selective}.}
Although not part of our analysis (but included in the uncompensated curve of Fig.~\ref{fig:KeyRateNR}), we  attempted to remove these trapping sets in our codes by using the algorithm of \cite{tian2004selective} so that fewer iterations will be used \cite{nguyen2012construction}.  This reduced the number of decoding iterations by approximately $15\%$ for $N_R = 10^6$ but did not remove the trapping sets completely. The remnant trapping sets
inside the LDPC matrices lead to a larger number of decoding iterations than predicted by $D_j$. Determined by the Density Evolution Algorithm, $D_j$ is a lower bound due to the assumption of cycle-free matrices and infinitely long block length\cite{landner2005algorithmic}.
%For example, we find that  the number of decoding iterations is $14\%$ higher than $D_j$ for $N_R=10^6$.
Secondly, we note that selecting $R_j$ so that $\sum_{j=0}^{m-1}R_j$ achieves $C_{Finite}$ may lead to a $p_{Decode}$ higher than the given $\epsilon_{EC}$.
In our experiment (and included in the shown results), we find that the $K_{exp}^\prime$ is $9\%$ lower than the $K_{Finite}^\prime$ predicted by Eq.~\ref{eq:BPSRate} due to the gap between $\sum_{j=0}^{m-1}R_j$ and $C_{Finite}$.

\subsection{Final-key Effect for a Given $N_o$}
\label{section:KeyRateFiniteResult}
%In the finite-key analysis, setting a large $N_e$ is to reduce the length of the confidence intervals when estimating $T$ and $\xi_{ch}$ and obtain a tight upper bound of the estimated $\chi_{BE}$. However, for a given $N_o$, setting $N_e$ closer to $N_o$ leads to a reduction on $K$ because fewer quantum signals are used for key extraction.
In the satellite-based scenario, Alice and Bob starts the protocol with only $N_o$ quantum signals because the satellite is only visible to the ground station for a limited time frame. In this section, we revisit the analysis of the final key rate in the finite-key regime and conduct a numerical search to show how the final key rate $K$ is affected by $N_e$, for a given $N_o$ and $\epsilon$.

\begin{figure}[h]
	\centering
	\includegraphics[width=0.5\textwidth]{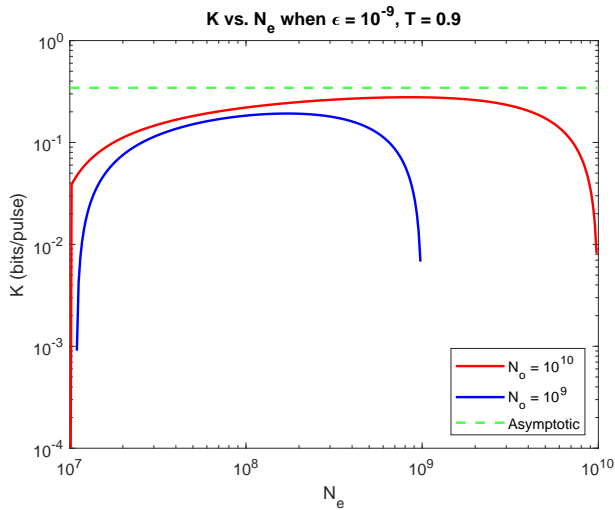}
	\caption{$K$ (in bits per pulse) vs. $N_e$ when $N_o = 10^9$ (blue) and $N_o = 10^{10}$ (red). Here, we adopt the standard CV-QKD settings except that $N = 2\left(N_o - N_e\right)$ varies for different $N_e$.\label{fig:KvsNeGivenNo}}
\end{figure}

%In Fig.~\ref{fig:KvsNeGivenNo}, we observe that $K$ is cut off when $N_e$ approaches to $10^7$ or $10^9$ (for $N_o = 10^9$). This observation can be explained as follows. The key rate $K$ is cut off at $10^7$ because $N_e$ is not sufficient to reduce the length of confidence intervals when estimating $T$ and $\xi_{ch}$ to result in a positive $K$. Setting $N_e$ close to $N_o$ also reduces $K$ because the number of quantum signals left for reconciliation and privacy amplification is close to zero. The explanation can also be applied for $N_o=10^{10}$. In Fig.~\ref{fig:KvsNeGivenNo}, we also find that, although $N_e$ requires further optimisation to maximise $K$, setting $N_e = \frac{N_o}{2}$ is an acceptable compromise between the reducing the finite-key effect of $K$ and preserving quantum signals for the classical post-processing. 
In Fig.~\ref{fig:KvsNeGivenNo}, we observe that $K$ is cut off when $N_e$ approaches $10^7$ and $10^9$ (for $N_o = 10^9$). At $N_e=10^7$, the parameter confidence intervals are not consistent with a positive $K$. As $N_e$ approaches $N_o$, $K$ decreases rapidly since the number of quantum signals for reconciliation approaches zero. Similar remarks can also be applied for $N_o = 10^{10}$. In Fig.~\ref{fig:KvsNeGivenNo}, we see that setting $N_e = \frac{N_o}{2}$ is an acceptable compromise between accommodating finite-key effects and preserving enough quantum signals for the post-processing. In the appendix, we investigate varying $N_e$ but where it is always constrained to $N_e=\frac{N_o}{2}$.

\section{Discussion}
We close our work with a brief discussion on  recent  developments in high-rate CV-QKD reconciliation   via the massive parallelisation offered by GPUs and FPGAs.
%There were major breakthroughs of achieving a higher reconciliation rate for CV-QKD systems in recent years due to the use of GPU or FPGA.
In \cite{milicevic2017key}, a GPU-based LDPC decoder was implemented, achieving a rate of $9\times 10^6$ bits per second. In this implementation  all GPU threads were used to minimise the decoding time of a \emph{single} LDPC block of $2^{20}$ bits. In \cite{wang2018high} and \cite{li2020high}, the reconciliation rates were further increased to $3\times10^7$ and $6\times 10^7$ bits per second, respectively, by simultaneously decoding \emph{multiple} LDPC blocks of length $10^6$ bits on a GPU. To our knowledge, the highest reconciliation rate obtained thus far is $2\times10^8$ bits per second  - an outcome based on an FPGA \cite{yang2020high}. All of these works show promise for the delivery of practical CV-QKD systems in which reconciliation does not become the bottleneck of the QKD process. However, none have introduced the type  of optimisation we have introduced in this work and,  therefore, all are likely candidates for further improvement in terms of the choice of the optimal block length. Based on our results we would anticipate this improvement to be significant for a wide range of CV-QKD parameter settings. Our work is also different from the above works in the following (less important) aspects.

%\textit{1) The LDPC block length considered.} The LDPC block lengths considered in the works above are mostly chosen as $10^5$ - $10^6$ bits, generating the highest reconciliation rate on the specific hardware. However, as shown in this work, the LDPC block lengths are not optimal  and further optimisation on the LDPC block length is still possible.

\textit{1) Reconciliation schemes for satellite-based CV-QKD.} High-speed implementations realised in \cite{wang2018high,li2020high} have used multidimensional reconciliation \cite{leverrier2008multidimensional}. This multidimensional scheme is preferred for low SNR - but not so for the higher SNRs available via the Gaussian post-selection technique  - a technique likely to be more useful in the satellite context \cite{hosseinidehaj2018satellite}.

% Our work shows that SR is also a viable candidate for satellite-based CV-QKD due to the application of Gaussian Post-selection\cite{hosseinidehaj2018satellite} technique.

\textit{2) Low probability of reconciliation failures.}
In CV-QKD, Alice and Bob have to discard a block of reconciled bits if they detect a reconciliation failure (coding error) for that block. To compensate for the discarded bits, additional quantum signals need to be transmitted and reconciled, causing unwanted delays. Such delays can be problematic for satellite-based systems since the satellite is not always visible to the ground station.
The FPGA-based reconciliation of \cite{yang2020high} may suffer from this problem due to limited precision of arithmetic operations leading to higher reconciliation failures. As shown in many GPU-based works (including this work), GPU-based reconciliation offers less probability of reconciliation failure.

\section{Conclusion}
In this work, we have carried out a full-blown analysis and experimental implementation of a Slice Reconciliation scheme applied to a specific CV QKD protocol (with post-selection) under simulated channel conditions anticipated for satellite-to-Earth channels. We have provided the optimal solution for the classical reconciliation process for this CV-QKD protocol in the context of massive parallelisation under the finite key regime. More specifically, we have identified the optimal block length when a large-code block is to be subdivided so as to improve the final secure key rate in bits per second. Although our results were based on a specific CV-QKD protocol and a specific GPU architecture, the type of analysis we have introduced here will apply in general terms a large suite of CV-QKD protocols run over any form of architecture that offers massive parallelisation. Our results, therefore, pave the way to optimal reconciliation system design for a wide range of practical CV-QKD systems that operate in the finite key regime. As the demand on the finite key size grows (better security thresholds), and technology advances lead to larger quantum signalling rates, the importance of optimised multithreaded CV-QKD reconciliation will grow.

%\section*{Acknowledgment}

%This work was supported by the Research Training Program (RTP) Fee Offset and the Postgraduate Research Support Scheme (PRSS) at the University of New South Wales.



%\bibliographystyle{IEEEtran}
 \bibliographystyle{mybst}
%\bibliography{IEEEabrv,teleportation}

\bibliography{CV_QKD}
\appendix
In this appendix, we elaborate on the estimation of channel parameters, $T$ and $\xi_{ch}$, from $N_e$ quantum signals and determination of the upper bound of $S_{BE}^{\epsilon_{PE}}$ based on the estimated $T$ and $\xi_{ch}$ for a given $N$. Here, we closely follow the methodology in \cite{kish2020feasibility} (and references therein).

The parameter estimation at Step 4 of our protocol is a two-step process. Firstly, Alice and Bob estimate each coefficient in the covariance matrix between the shared states based on $N_e$ (randomly selected) quantum signals sent from Bob. Then Alice uses these estimated coefficients to determine $T$ and $\xi_{ch}$. In the asymptotic regime, the estimation of $T$ and $\xi_{ch}$ is exact since Alice and Bob use an infinite number of quantum signals. The following the two functions will be useful,
\begin{eqnarray}
	&F_1(v_1,v_2) = \sqrt{\frac{v_1 + \sqrt{v_1^2 - 4v_2}}{2}}\,,\label{eq:F1}\\
	&F_2(v_1,v_2) = \sqrt{\frac{v_1 - \sqrt{v_1^2 - 4v_2}}{2}}\,.\label{eq:F2}
\end{eqnarray}
Alice can determine the Holevo Information between Bob and Eve's states $\chi_{EB}$ via\cite{grosshans2005collective,navascues2006optimality,fossier2009improvement}
\begin{equation}
\label{eq:chi_EB}
\chi_{EB} = \chi_E - \chi_{E|B}\,,
\end{equation}
where $\chi_E$ is Eve's von Neumann Entropy before Bob makes his heterodyne detection and $\chi_{E|B}$ is Eve's von Neumann Entropy after his detection. The term $\chi_E$ is given by
\begin{equation}
\label{eq:chi_E}
\chi_E = Z\left(\frac{\psi_1 - 1}{2}\right) +Z\left(\frac{\psi_2 - 1}{2}\right)\,,
\end{equation}
where
\begin{equation}
	Z(z) = (z+1)\log{(z+1)} - z\log{z}\,.
\end{equation}
We define that $\psi_1=F_1(\Psi_1,\Psi_2)$ and $\psi_2=F_2(\Psi_1,\Psi_2)$ to be the symplectic eigenvalues of the covariance matrix of the shared states (before Bob's heterodyne detection) where
\begin{eqnarray}
%	\psi_1 = \sqrt{\frac{\Psi_1 + \sqrt{\Psi_1^2 - 4\Psi_2}}{2}}\,,\\
%	\psi_2 = \sqrt{\frac{\Psi_1 - \sqrt{\Psi_1^2 - 4\Psi_2}}{2}}\,,
&\Psi_1 = (V_A + 1)^2(1-2T) + 2T + T^2\left(V_A+1+\chi_{ch}\right)\,,\label{eq:Psy1}\\	
&\Psi_2 = T^2\left((V_A+1)\xi_{ch} + 1  \right)\,,\label{eq:Psy2}\\
&\chi_{ch} = \frac{1-T}{T}+\xi_{ch}\,. \label{eq:chi_chi}
\end{eqnarray}
The term $\chi_{E|B}$ is given by
\begin{equation}
\label{eq:chi_EaB}
\chi_{E|B} = Z\left(\frac{\theta_1 - 1}{2}\right) +Z\left(\frac{\theta_2 - 1}{2}\right) +Z\left(\frac{\theta_3 - 1}{2}\right) \,,
\end{equation}
where $\theta_1$, $\theta_2$ and $\theta_3$ are the symplectic eigenvalues of the covariance matrix of the shared states (after Bob's heterodyne detection). Specifically, we have $\theta_1 = F_1\left(\Theta_1,\Theta_2\right)$ and $\theta_2 = F_2\left(\Theta_1,\Theta_2\right)$ where
\begin{eqnarray}
&\begin{aligned}
\Theta_1 =& \bigg(\Psi_1\chi^2_d + \Psi_2+1 \bigg.\\
&\bigg. + 2\chi_d\left( T\left(V_A+1+\chi_{ch}\right)+\left(V_A+1\right)\sqrt{\Psi_2}\right) \bigg.\label{eq:Theta1}\\
&\bigg. + 2T\left(V_A^2+2V_A\right)\bigg)\frac{1}{T^2\left(V_A+1+\chi\right)}\,,\\
\end{aligned}\\	
&\Theta_2 = \left(\frac{V_A+1+\chi_d\sqrt{\Psi_2}}{T\left(V_A+1+\chi\right)}\right)^2\,,\label{eq:Theta2}\\
&\chi_d = \frac{2-\eta_d}{\eta_d}+\frac{2\chi_d}{\eta_d}\,,\label{eq:chi_d}\\
&\chi = \chi_{ch} + \frac{\chi_d}{T}\,,\label{eq:chi}
\end{eqnarray}
where $\eta_d$ is the detection efficiency and we set $\eta_d=1$ for simplicity. It is known that $\theta_3=1$ under the assumption of Gaussian collective attack\cite{fossier2009improvement}. Therefore, we have $Z\left(\frac{\theta_3 - 1}{2}\right) =0$.
%We note that the reason why two symplectic eigenvalues are found when calculating $\chi_E$ is that covariance matrix of the shared states involves only Alice and Bob's systems. While, three symplectic eigenvalues are found when calculating $\chi_{E|B}$ because covariance matrix of the shared states involves Alice, Bob and Eve's systems.

However, the estimation of $T$ and $\xi_{ch}$ is not exact in the finite-key regime. The estimated $T$ and $\xi_{ch}$ are subject to statistical fluctuations that leads to a deviation of the estimated $T$ and $\xi_{ch}$ from their true values (since Alice and Bob use only $N_e$ signals for the estimation at Step 4). The impact of using a finite number of quantum signals for parameter estimation in the security analysis is twofold. Firstly, the protocol will fail with a probability of $\epsilon_{PE}$ if the true value of $T$ or $\xi_{ch}$ is out of the confidence interval set by that $\epsilon_{PE}$. Secondly, the amount of the deviation of the estimated $T$ and $\xi_{ch}$ from their true values is probabilistic. The lower and upper limits of the confidence interval of the estimated $T$ for a given $\epsilon_{PE}$ are given by\cite{leverrier2010finite,jouguet2012analysis}
\begin{eqnarray}
T^L = \left(\hat{t} - \tau_{\epsilon_{PE}/2}\sqrt{\frac{\hat{\sigma}^2}{N_e V_A}}\right)^2\label{eq:estT}\,,\\
T^U = \left(\hat{t} + \tau_{\epsilon_{PE}/2}\sqrt{\frac{\hat{\sigma}^2}{N_e V_A}}\right)^2\label{eq:estT2}\,,
\end{eqnarray}
where $\tau_{\epsilon_{PE}/2} = Q^{-1}(\frac{\epsilon_{PE}}{2})$; and $\hat{t}$ and $\hat{\sigma}$ are the estimators for $T$ and $\xi_{ch}$, respectively. Similarly, the lower and upper limits of the confidence interval of the estimated $\xi_{ch}$ for a given $\epsilon_{PE}$ are given by\cite{leverrier2010finite,jouguet2012analysis}
\begin{eqnarray}
\xi_{ch}^L = \frac{\hat{\sigma}^2 - \tau_{\epsilon_{PE}/2}\frac{\hat{\sigma}^2\sqrt{2}}{\sqrt{N_e}}+1+\xi_d}{\hat{t}^2}\,,\label{eq:estXi2}\\
\xi_{ch}^U = \frac{\hat{\sigma}^2 + \tau_{\epsilon_{PE}/2}\frac{\hat{\sigma}^2\sqrt{2}}{\sqrt{N_e}}-1-\xi_d}{\hat{t}^2}\,,\label{eq:estXi}
\end{eqnarray}
respectively.

Based on the above, we can now determine $S_{BE}^{\epsilon_{PE}}$, i.e. the upper bound of $\chi_{BE}$ in the finite-key regime. Firstly, for the purpose of analysis, we set the expectation of $\hat{t}$ and $\hat{\sigma}$ as $\sqrt{\eta_d T}$ and $T\eta_d \xi_{ch}+1+\xi_d$, repectively. Then, we replace $T$ and $\xi_{ch}$ in Eqs.~\ref{eq:Psy1} to~\ref{eq:chi_chi} and Eqs.~\ref{eq:Theta1} to~\ref{eq:chi}  with $T^L$ and $\xi_{ch}^U$, respectively. Next, we determine $S_{BE}^{\epsilon_{PE}}$ by using Eqs.~\ref{eq:F1},~\ref{eq:F2} to determine all the symplectic eigenvalues. Finally, we use Eq.~\ref{eq:chi_E},~\ref{eq:chi_EaB} and~\ref{eq:chi_EB} to obtain $S_{BE}^{\epsilon_{PE}}$.

\begin{figure}[!htbp]
	\centering
		\includegraphics[width=0.45\textwidth]{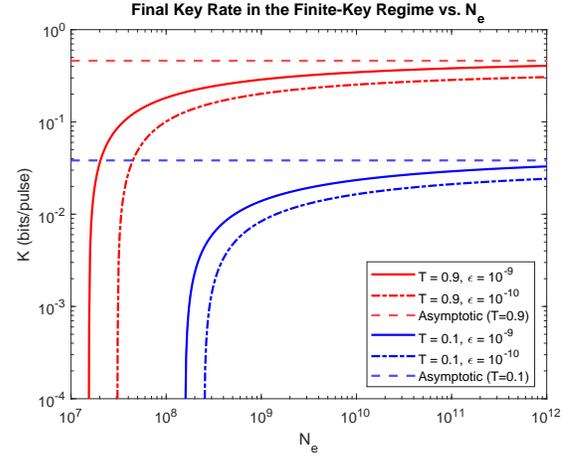}
		\caption{$K$ (in bits per pulse) vs. $N_e$. Here we adopt the standard CV-QKD setting except for $N_o$ for all the curves. For all the curves, we assume $N_e = \frac{N_o}{2}$. \label{fig:KvsNe}}		
\end{figure}
The motivation of setting a large $N_e$ is to reduce the length of the confidence intervals when estimating $T$ and $\xi_{ch}$. In Fig.~\ref{fig:KvsNe}, we compare the impact on $K$ when setting different $N_e$. For all the curves in Fig.~\ref{fig:KvsNe}, we assume $N_e = \frac{N_o}{2}$ (see Section~\ref{section:KeyRateFiniteResult} for the case of varying $N_e$ for a given $N_o$). The ``take-away'' message is that, for a given $\epsilon$, setting a large $N_e$ is necessary for most CV-QKD deployments if a significant reduction of $K$ is to be avoided.
%\includegraphics[width=0.5\textwidth]{KvsNe}

%The motivation of setting a large $N_e$ is to reduce the length of the confidence intervals when estimating $T$ and $\xi_{ch}$. In Fig.~\ref{fig:KvsNe}, we compare the impact on $K$ when setting different $N_e$. For all the curves in Fig.~\ref{fig:KvsNe}, we assume $N_e = \frac{N_o}{2}$ (see Section~\ref{section:KeyRateFiniteResult} for the case of varying $N_e$ for a given $N_o$). The ``take-away'' message is that, for a given $\epsilon$, setting a large $N_e$ to avoid a significant reduction of $K$ is necessary for most CV-QKD deployments.

\end{document}